\documentclass[structabstract]{aa}
\usepackage{graphicx}
\usepackage{longtable}
\usepackage[authoryear]{natbib}
\bibpunct{(}{)}{;}{a}{}{,}
\usepackage{txfonts}
\begin{document}

\title{Astrophysical parameters of the peculiar X-ray transient IGR~J11215$-$5952\fnmsep\thanks{Based on observations collected
   at the European Southern Observatory, La Silla, Chile (ESO 078.D-0172)}}
\author{J.~Lorenzo\inst{1}
\and I.~Negueruela\inst{1}
\and N.~Castro\inst{2}
\and A.J.~Norton\inst{3}
\and F.~Vilardell\inst{1,4}
\and A.~Herrero\inst{5,6}}

\institute{
Departamento de F\'{i}sica, Ingenier\'{i}a de Sistemas y
  Teor\'{i}a de la Se\~{n}al, Universidad de Alicante, Apdo. 99, E03080
  Alicante, Spain\\
\email{ignacio.negueruela@ua.es}
 \and Argelander Institut f\"ur Astronomie, Auf den H\"ugel 71, Bonn, 53121, Germany
\and
Department of Physical Sciences, The Open
University, Walton Hall, Milton Keynes, MK7 6AA, UK
\and Institut d'Estudis Espacials de Catalunya, Edifici Nexus, c/ Capit\`{a}, 2-4, desp. 201, E-08034 Barcelona, Spain
\and Instituto de Astrof\'{i}sica de Canarias, 38200 La Laguna, Tenerife, Spain
\and Departamento de Astrof\'{\i}sica, Universidad de La Laguna, Avda. Astrof\'{\i}sico Francisco S\'anchez, s/n, 38071 La Laguna, Spain}

\titlerunning{The peculiar X-ray transient IGR~J11215$-$5952}

\abstract{The current generation of X-ray satellites has discovered many new X-ray sources that are difficult to classify within the well-described subclasses. The hard X-ray source IGR~J11215$-$5952 is a peculiar transient, displaying very short X-ray outbursts every 165 days.}
{To characterise the source, we obtained high-resolution spectra of the optical counterpart, HD~306414, at different epochs, spanning a total of three months, before and around the 2007 February outburst with the combined aims of deriving its astrophysical parameters and searching for orbital modulation.}
{We fit model atmospheres generated with the \textsc{fastwind} code to the spectrum, and used the interstellar lines in the spectrum to estimate its distance. We also cross-correlated each individual spectrum to the best-fit model to derive radial velocities.}
{From its spectral features, we classify HD~306414 as B0.5\,Ia. From the model fit, we find $T_{{\rm eff}}\approx  24\,700\:{\rm K}$ and $\log g\approx2.7$, in good agreement with the morphological classification. Using the interstellar lines in its spectrum, we estimate a distance to HD~306414 $d\ga7\:$kpc. Assuming this distance, we derive $R_{*}\approx40\:R_{\sun}$ and $M_{{\rm spect}}\approx30\:M_{\sun}$ (consistent, within errors, with $M_{{\rm evol}}\approx38\:M_{\sun}$, and in good agreement with calibrations for the spectral type). Analysis of the radial velocity curve reveals that radial velocity changes are not dominated by the orbital motion, and provide an upper limit on the semi-amplitude for the optical component $K_{{\rm opt}}\la11\pm6\:{\rm km}\,{\rm s}^{-1}$. Large variations in the depth and shape of photospheric lines suggest the presence of strong pulsations, which may be the main cause of the radial velocity changes. Very significant variations, uncorrelated with those of the photospheric lines  are seen in the shape and position of the H$\alpha$ emission feature around the time of the X-ray outburst, but large excursions are also observed at other times.}
{HD~306414 is a normal B0.5\,Ia supergiant. Its radial velocity curve is dominated by an effect that is different from binary motion, and is most likely stellar pulsations. The data available suggest that the X-ray outbursts are caused by the close passage of the neutron star in a very eccentric orbit, perhaps leading to localised mass outflow.}

\keywords{stars: binaries: close - stars: evolution - stars: individual: HD 306414  - stars: pulsars - stars: supergiants - X-rays: stars - X-rays: individual: IGR J11215$-$5952}

\maketitle

\section{Introduction}

The hard X-ray transient IGR~J11215$-$5952 was discovered by the {\it INTEGRAL} satellite \citep{lubin2005}. A {\it Swift} localisation confirmed its unambiguous identification as the bright B-type supergiant HD~306414 \citep{rom08}. IGR~J11215$-$5952 has been observed to present a number of short X-ray outbursts, separated by long intervals of quiescence, with a recurrence time of 164.6~d \citep{rom09}. During each outburst, the source is detected for $\sim8$~d. The outbursts appear to have a recurrent structure, consisting of a fast rise, a bright peak (reaching $L_{{\rm X}}\approx10^{36}\:{{\rm erg}}\,{{\rm s}}^{-1}$ for the assumed distance of 6.2~kpc) that may last only $\sim1$~d each cycle, and a gradual decay. However, the X-ray emission during the decay seems to consist of many flares superimposed and the sparse coverage does not permit ruling out that some of them reach similar luminosity to the peak \citep{rom09}. During the flares, pulsations at $P_{{\rm S}}=187\:{\rm s}$ are seen, indicating that the compact object is a magnetised neutron star. The X-ray spectra are typical of a high-mass X-ray binary.

 Outside the outbursts, the X-ray luminosity has been observed to be below the detection limit of {\it Swift}, roughly corresponding to $L_{{\rm X}}\la5\times10^{33}\:{{\rm erg}}\,{{\rm s}}^{-1}$ at the distance assumed \citep{rom08}. Seven outbursts were reported between 2003 and 2008, all happening at the recurrence time. No detections outside these outbursts are reported.

 The counterpart, HD~306414, has not been studied in detail. It was classified as a B1\,Ia supergiant \citep{vijap1993}, and the scarce photometry available in the literature does not suggest strong photometric variability \citep{negue2005}. The X-ray behaviour of the source, characterised by irregular flaring, seems typical of neutron stars accreting from the wind of a supergiant. Because of the regular outbursts and lack of detection outside them,  IGR~J11215$-$5952 has been classed as a supergiant fast X-ray transient \citep[SFXT;][]{rom09}. A few of these objects show similar behaviour, displaying periodic outbursts \citep[e.g. IGR~J18483$-$0311;][]{romano10}, though others seem to flare at irregular intervals, even if some orbital modulation of the X-ray emission is present. The recurrence of outbursts at fixed intervals is widely interpreted as a signature of the orbital period, while the lack of X-ray emission outside these outbursts argues for an eccentric orbit.

In this paper, we try to improve our understanding of IGR~J11215$-$5952 by studying in detail its optical counterpart, HD~306414. In Sect.~2, we present our observational dataset. In Sect.~3, we fit the spectrum with a model atmosphere to determine the stellar parameters. In Sect.~4, we discuss the radial velocity variations seen in the spectrum. Finally, in Sects.~5 and~6, we discuss the implications of these results and draw some conclusions.

\section{Observations  and description of the spectrum}

Spectra of HD 306414 were obtained with the \textsc{feros} instrument mounted on the ESO/MPG 2.2~m telescope, located at the European Southern Observatory (ESO) in La Silla (Chile). Observations were obtained on eleven dates irregularly spaced between December 2006 and February 2007. Two additional \textsc{feros} spectra were taken in March 2009, with another one taken in May 2009 using the same instrumentation. The 14 spectra were bias subtracted, flatfield corrected, extracted using the optimum mode and wavelength calibrated using the standard \textsc{feros} Midas pipeline. The spectra are characterised by a very wide spectral coverage, going from $\sim$3500\,\AA\ to $\sim$9200\,\AA, and a resolving power $R\sim48\,000$. The 39 \'echelle orders are blaze-corrected and merged into a single spectrum by the pipeline.
The complete log of observations is given in Table~\ref{table:log}.

\begin{figure*}[ht]
\resizebox{\textwidth}{!}{
\includegraphics[]{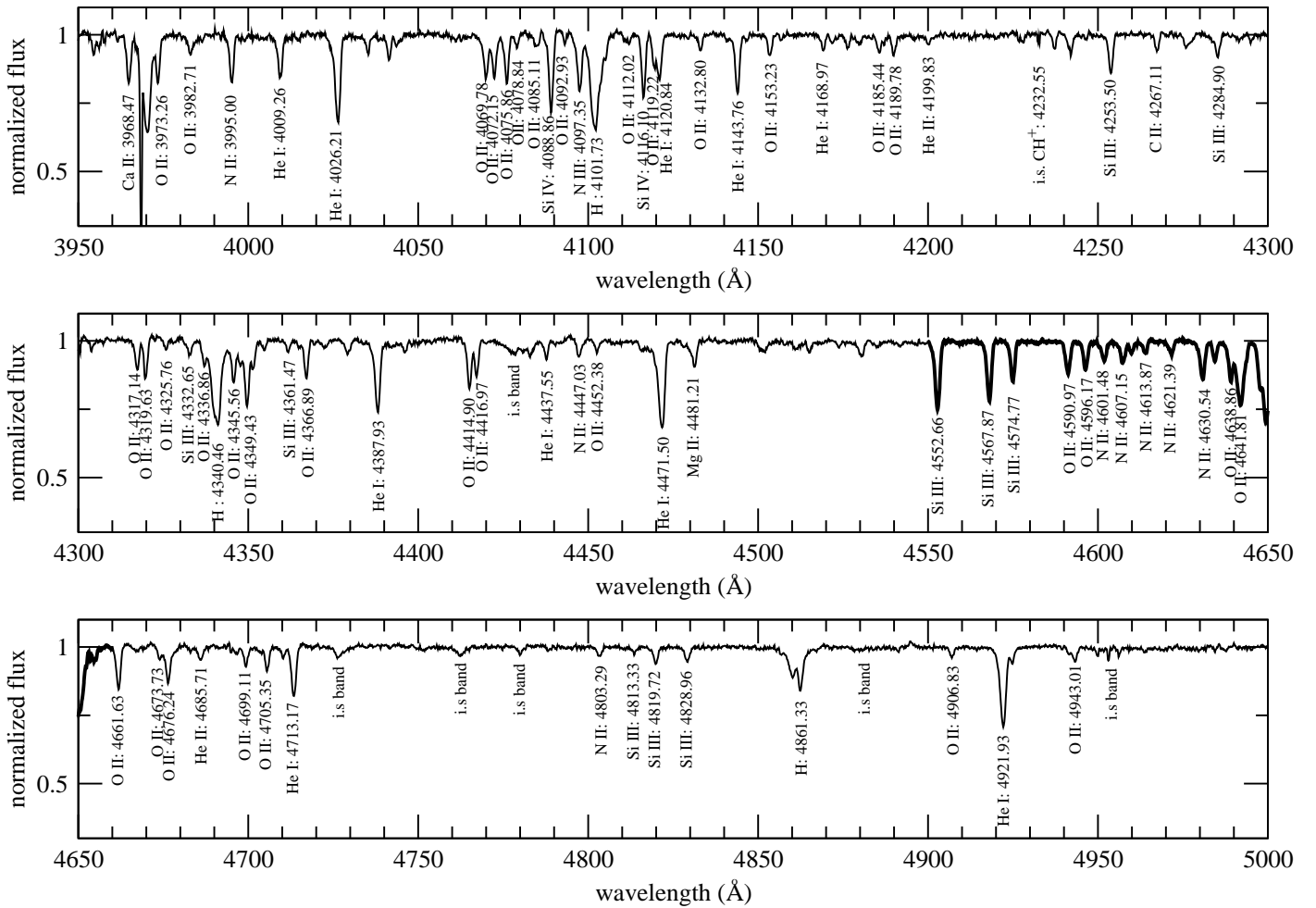}}
\caption{Spectrum of HD~306414 covering from 3950\,\AA~till 5000\,\AA. We show identifications for Balmer and \ion{He}{i}, as well as the strongest lines of Si\,{\sc iii}, Si\,{\sc iv}, O\,{\sc ii}, N\,{\sc ii} and N\,{\sc iii}. The spectral region selected to determine radial velocities via cross-correlation is highlighted. \label{spec}}
\end{figure*}

The classification region of a typical spectrum is shown in Fig.~\ref{spec}, corresponding to the Dec 22, 2006 observation, with labels providing line identification. The remainder of the spectrum is not shown (except for selected regions in Figures~\ref{feiii}, \ref{nalines} and~\ref{alfa}), as there are very few stellar features present, and they are as expected for the spectral type derived below  (see \citealt{negue2010} for a description of typical features). The spectra only show variability in the H$\alpha$ and H$\beta$ lines. H$\alpha$ is always in emission, but the line shows variations both in shape and centroid position (see Sect.~\ref{sec:radial} ). H$\beta$ is regularly in absorption, but occasionally shows emission components, as in the spectrum shown in Fig.~\ref{spec}. The absorption spectral features correspond to an early B supergiant, with \ion{H}{i} and \ion{He}{i} lines showing similar intensities and many narrow lines of ionised metals, most notably \ion{O}{ii}, \ion{N}{ii} and \ion{Si}{iii}/{\sc iv}.

\begin{table}
\caption{Log of observations of high-resolution spectra sorted by date.\label{table:log}}
\smallskip
\begin{center}
\begin{tabular}{lcc}
\hline
\hline
\noalign{\smallskip}
Number&Date&MJD\\
\noalign{\smallskip}
\hline
\noalign{\smallskip}
1 &03/12/2006 7:52:00&54072.33\\
2 &11/12/2006 5:41:00&54080.24\\
3 &15/12/2006 6:34:00&54084.27\\
4 &22/12/2006 8:38:00&54091.36\\
5 &29/12/2006 6:22:00&54098.27\\
6 &17/01/2007 3:30:00&54117.15\\
7 &05/02/2007 8:50:00&54136.37\\
8 &11/02/2007 4:39:00&54142.19\\
9 &13/02/2007 1:38:00&54144.07\\
10 &16/02/2007 2:22:00&54147.10\\
11&16/02/2007 2:40:00&54147.11\\
12&19/03/2009 7:41:28&54909.32\\
13&21/03/2009 3:27:10&54911.14\\
14&13/05/2009 2:09:00&54964.09\\
\noalign{\smallskip}
\hline
\end{tabular}
\end{center}
\end{table}

\section{Results}

\subsection{Spectral classification}

HD~306414 has previously been classified as B1\,Ia. For early B stars, the main temperature diagnostic is the ratio between the \ion{Si}{iii}~4553\,\AA\ and \ion{Si}{iv}~4089\,\AA\ lines. In our case, the relative strengths are very similar, placing HD~306414 between spectral types B0.5 and B0.7. The \ion{He}{ii}~4686\,\AA\ line is very weakly present, and \ion{He}{ii}~4200 \& 4541\,\AA\ are hardly detectable, corroborating that the star lies in this spectral range. Luminosity sensitive ratios, like that of   \ion{Si}{iv}~4116\,\AA\ to \ion{He}{i}~4121\,\AA, support a high luminosity, with the intensity of the \ion{O}{ii} lines, and the ratio of \ion{Si}{iii}~4553\,\AA\ to \ion{He}{i}~4388\,\AA\ favouring a Ia class.

\begin{figure}[]
\resizebox{\columnwidth}{!}{\includegraphics[clip]{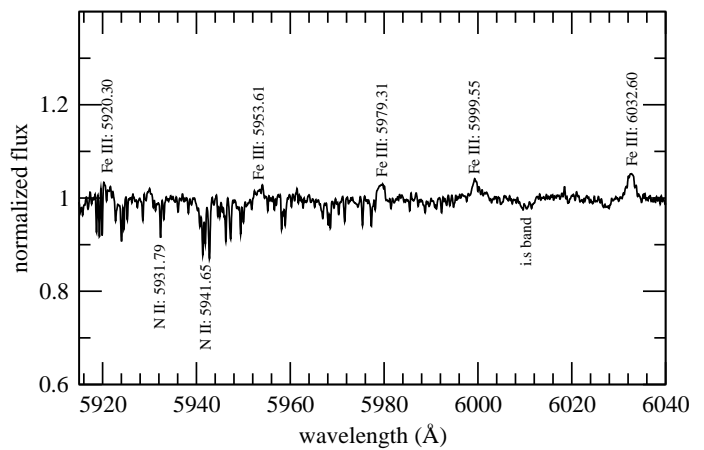}}
\caption{The Fe\,{\sc iii} lines of multiplets 115 and 117 in emission in the spectrum of HD~306414.\label{feiii}}
\end{figure}

Therefore we adopt a spectral type B0.5\,Ia, though noting that the star is slightly later than the standard, $\kappa$~Ori. Apart from H$\alpha$, the most remarkable spectral features outside the range shown are the \ion{Fe}{iii} lines of multiplets 115 and 117 in emission between 5920 and 6032\,\AA\ (see Fig.~\ref{feiii}). These lines are seen in emission in the spectra of very luminous supergiants. \citet{wolf1985} proposed that their presence in emission could be used to discriminate between normal B supergiants and B hypergiants. However, HD~306414 is not a hypergiant, as shown by the fact that \ion{He}{i}~4387\,\AA\ is stronger than \ion{Si}{iii}~4553\,\AA\ \citep[and Fig.~\ref{spec}]{walb90}. Indeed, we find the \ion{Fe}{iii} lines in emission in the spectra of several early-B supergiants of luminosity class Ia in the UVES POP database \citep{bagnulo}. Their presence in the spectrum of HD~306414 represents direct confirmation of its high luminosity.

\subsection{Interstellar lines and distance estimation}
We used the interstellar lines in the spectrum of HD~306414 to study the radial velocity distribution of interstellar material along its line of sight. We calculated the velocity scale with respect to the local standard of rest (LSR) by assuming that the Sun's motion with respect to the LSR corresponds to $+16.6\:{\rm km}\,{\rm s}^{-1}$ towards Galactic coordinates $l=53\degr$; $b=+25\degr$.

\begin{figure*}
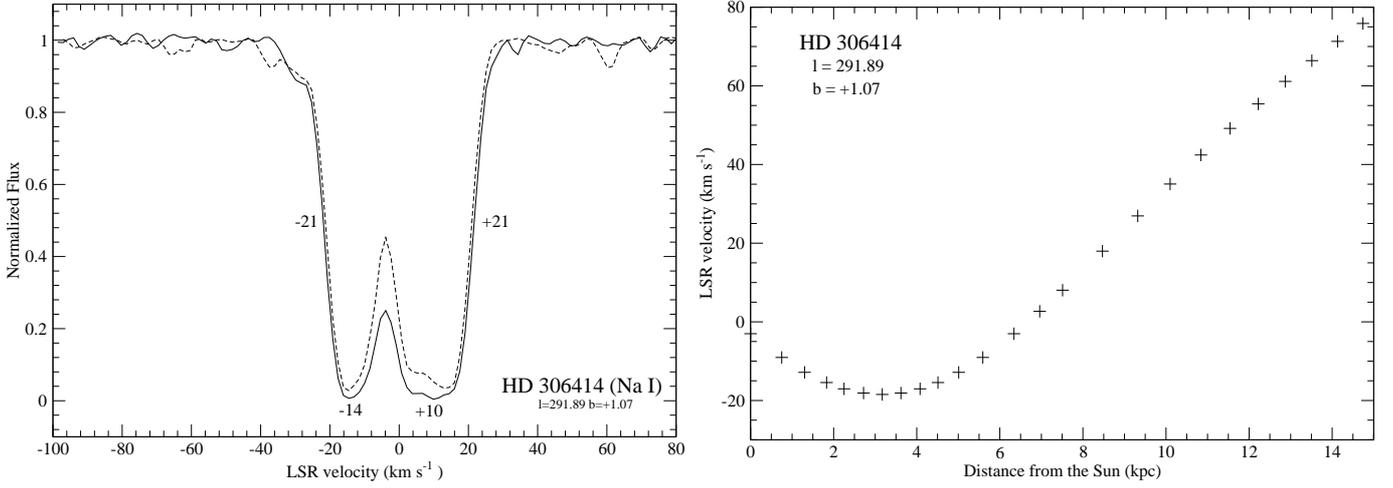

\resizebox{\columnwidth}{!}{
\includegraphics[clip]{NaHD306414.eps}}
\resizebox{\columnwidth}{!}{
\includegraphics[clip]{lrssun306414}}
\caption{{\bf Left panel: }Interstellar Na\,{\sc i} D doublet (5889.953 \,\AA$\:$ full line, 5895.923 \,\AA$\:$ dashed line) of HD 306414. {\bf Right panel: } Radial velocity with respect to the local standard of rest (LSR) due to Galactic rotation as a function of distance.\label{nalines}}
\end{figure*}

In Fig.~\ref{nalines}, we show the interstellar Na\,{\sc i} D lines; both present identical morphologies, with two well-separated components. Other interstellar lines, such as the \ion{Ca}{ii} K \& H doublet, have an almost identical structure. The \ion{K}{i}~7699\,\AA\ line, which is not saturated, has very similar edge velocities. In the Na\,{\sc i} D lines, the wider component is centred on $+10\:{\rm km}\,{\rm s}^{-1}$ and the narrower component is centred on $-14\:{\rm km}\,{\rm s}^{-1}$. The total line width at half height is $\approx42\:{\rm km}\,{\rm s}^{-1}$. Both lines have the same wing profiles, but the D2 line is more saturated than D1. In Fig.~\ref{nalines}, we also show the Galactic rotation curve in the direction to HD~306414 ($l=291\fdg89$, $b=+1\fdg07$), computed assuming circular Galactic rotation and adopting the rotation curve of \citet{br93}, with a circular rotation velocity at the position of the Sun ($d_{{\rm GC}}=8.5$~kpc) of $220\:{\rm km}\,{\rm s}^{-1}$. Along this line of sight, LSR velocities start at small negative values and become more negative with distance until reaching a minimum at $-18.5\:{\rm km}\,{\rm s}^{-1}$ (at a Galactocentric distance of 7.9~kpc, 3.2~kpc away from the Sun). From then on, radial velocities increase with distance, becoming positive at a distance of 6.5~kpc from the Sun.

The line of sight in the direction to HD~306414 passes first through the Southern Coalsack (distance about 170~pc), one of the most prominent dark nebulae in the southern Milky Way \citep{ny98}, which is responsible for the strong absorption at low (negative) velocities. It then follows along the Sagittarius-Carina spiral arm, which is seen almost tangentially. The first intersection with the arm takes place between $\sim$1.0 and $2.5$~kpc. The longitude-velocity diagram of CO emission integrated over latitude \citep{co85} shows CO emission with negative velocities, reaching $-45\:{\rm km}\,{\rm s}^{-1}$, which do not agree with the model prediction. The interstellar lines in HD~306414 show components at similarly high negative velocities (with the half-depth edge at $-21\:{\rm km}\,{\rm s}^{-1}$ and a component reaching almost  $-35\:{\rm km}\,{\rm s}^{-1}$), indicating that the star is located at a larger distance. In particular, the star must be more distant than two large molecular clouds (288.5+1.5 and 291.5$-$0.8) at a distance $d\approx3.3$~kpc \citep{co85}.

The far intersection with the Sagittarius-Carina arm (outside the solar circle) results in CO emission at positive LSR velocities \citep{co85}. The star HD~97253, located at $d\approx2.5$~kpc ($290\fdg8$, $+0\fdg1$), only shows negative radial velocities in its interstellar \ion{Na}{i} lines \citep[which are otherwise very similar to those of HD~306414; cf.][]{hunter06}, suggesting that the positive components in the interstellar spectrum of HD~306414 are not produced by clouds very close to the Sun\fnmsep\footnote{Other nearby stars have similar interstellar lines. HD~97534 ($290\fdg98$, $+0\fdg24$), located at about 3.9~kpc \citep{da91}, only displays negative velocities. HD~94910 (AG~Car; $289\fdg2$, $-0\fdg70$), located at a distance $\sim6$~kpc \citep{ho92,hu89} also shows very similar interstellar lines, but the long wavelength edge is affected by the presence of stellar emission components.}. Though \citet{co85} do not identify any large cloud producing absorption close to HD~306414, this absorption edge (reaching values as high as $+21\:{\rm km}\,{\rm s}^{-1}$) must be produced at large distances.
In view of this, we conclude that HD~306414 must be located on the near side of the second intersection with the Sagittarius-Carina arm, at a distance not less than, and perhaps slightly more than, $7$~kpc. The nearby massive open cluster NGC~3603 ($291\fdg5$, $-0\fdg4$) has a distance estimate of $7.6$~kpc \citep{melena08} and $v_{{\rm LSR}}= +14\:{\rm km}\,{\rm s}^{-1}$, in good agreement with the radial velocity curve model.

\subsection{Spectrum modelling}
\label{sec:spec}

To calculate the stellar parameters, we used the code \textsc{Fastwind} (an acronym for \textsc{F}ast \textsc{A}nalysis of \textsc{ST}ellar atmospheres with \textsc{WIND}s; \citealt{pu05,sa97}). \textsc{Fastwind} is a spherical non-LTE model atmosphere code with mass loss. The analysis is based on visual fitting of hydrogen Balmer and Si\,{\sc iii}/{\sc iv} lines \citep[see][for additional details]{castro12}. With the stellar parameters obtained from the analysis, we generated a synthetic spectrum, which will be the template used for the determination of radial velocities through the cross-correlation method.

We used as a reference the spectrum observed on December 15th, 2006.
As a first step we estimated the rotational velocity of the star, using the method of \citet{sergio}, as $v_{{\rm rot}}=50\:{\rm km}\,{\rm s}^{-1}$. With this $v_{{\rm rot}}$, a macroturbulence value $v_{{\rm mac}}=80\:{\rm km}\,{\rm s}^{-1}$ is needed to reproduce the profiles. The best fit is obtained for $T_{{\rm eff}}=24\,700$~K, $\log g=2.7$. The temperature is slightly cooler than obtained for other B0.5\,Ia supergiants \citep{markova,crowther}, but hotter than that of B0.7\,Ia stars \citep{crowther}, in good agreement with the spectral classification. The surface gravity is typical of the luminosity class.

The mass loss rate was determined from the fit to the H$\alpha$ line and must be considered approximate, as the line shows some variability. The value derived depends on $v_{\infty}$, which, in the absence of ultraviolet spectra, must be assumed to be the typical value for the spectral type. The wind was modelled without considering clumping. Model parameters are listed in Table~\ref{table:stellarpar}.

\begin{table}[!h]
\centering
\caption{Stellar parameters derived from the model fit (upper panel) and calculated using the photometry of \citet{klare} and assuming $d=7.0$~kpc (lower panel).\label{table:stellarpar}}
\begin{tabular}{lc}
\hline
\hline
\noalign{\smallskip}
$T_{{\rm eff}}~(10^{3}$K)&$24.2\pm1.0$\\
$\log g$ &$2.7\pm0.1$\\
$v\sin i$~(${\rm km}\,{\rm s}^{-1}$)&50\\
$U-B$&$-1.00$\\
$B-V$&$-0.20$\\
$V-R$&$-0.15$\\
\hline
\noalign{\smallskip}
$d$(kpc)&$7\pm1$\\
$M_{V}$&$-7.1\pm0.3$\\
$\log (L_{\ast}$/$L_{\sun}$)&5.$68\pm0.14$\\
$R_{\ast}/R_{\sun}$&$40\pm5$\\
$M_{\ast}/M_{\sun}$&$29\pm10$\\
$\log(\dot{M})$~($M_{\sun}\,{\rm yr}^{-1}$) & $-5.7\pm0.3$\\
$v_{\infty}$~(${\rm km}\,{\rm s}^{-1}$)&1230\fnmsep\tablefootmark{a}\\
\hline
\end{tabular}
\newline\\
\tablefoottext{a}{Assumed.}
\end{table}

The He relative abundance, $\epsilon=0.13\pm0.03$ is very slightly above solar, suggesting little chemical evolution. From the model fits, we derived chemical abundances for the species present in the spectrum. The values obtained are displayed in Table~\ref{table:abundances}. The C abundance is very inaccurate, because it is derived from only one line (\ion{C}{ii} 4267\,\AA), as \ion{C}{ii}~6578, 6582\,\AA\ are too weak at this spectral type. The Mg abundance was also obtained from one line (\ion{Mg}{ii}~4481\,\AA), but its value is consistent with the Si abundance. The N and O abundances were obtained by fitting a large number of lines. Not all the transitions are modelled with the same accuracy, but the values derived seem consistent. Based on them, the star seems to be slightly N-enhanced, suggesting little evolution, in agreement with the He abundance.

\begin{figure}[]
\resizebox{\columnwidth}{!}{\includegraphics[angle=+90,clip]{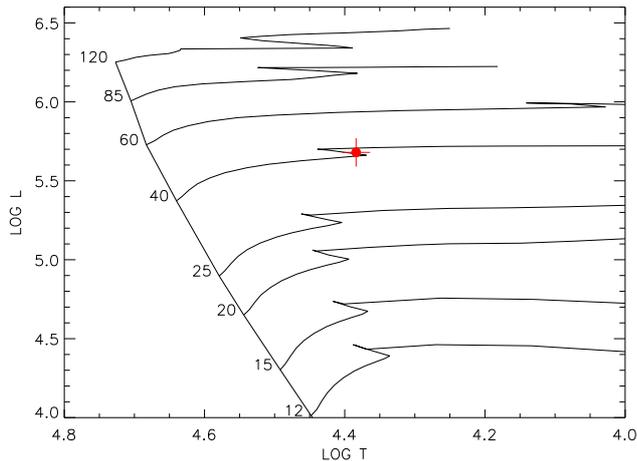}}
\caption{Hertzsprung-Russell diagram showing the evolutionary tracks
without rotation by \citet{scha1992}, and the position of HD~306414.
Numbers to the left of the ZAMS indicate initial mass in solar masses.\label{hr}}
\end{figure}

\subsection{Extinction and stellar parameters}

The combination of accurate stellar parameters
with a broad spectral energy distribution allows a good
determination of the extinction law and reddening. We used $UBV$ photometry from the literature \citep{klare}\fnmsep\footnote{Note that the $UBV$ photometry of \citet{drilling} shows excellent agreement with that of \citet{klare}.}, and $JHK_{{\rm S}}$ photometry from 2MASS \citep{skru06}. These data were used as input for the
$\chi^{2}$ code for parameterized modelling and characterisation of
photometry and spectroscopy {\sc chorizos}
implemented by \citet{maiz04}, which fits different extinction laws to the data and determines the values of $R$ and
$E(B-V)$ that produce best fits. From the fit, we obtain as most likely parameters, $R=4.2$, $A_{V}=3.00$, $E(B-V)=0.70$. The value of $R$ is rather high compared to the standard value $R=3.1$, but well within the range of values seen along different lines of sight.

Assuming a distance $d=7.0$~kpc, a lower limit based on the estimate in the previous section, the absolute magnitude for the source is $M_{V}=-7.1$. This value is within the usual range of magnitudes for early-B supergiants of luminosity class Ia \citep{crowther}. From this, we calculated
the absolute stellar parameters displayed in Table~\ref{table:stellarpar}. In order to assign formal errors to these values, we assumed conservative errors of $\pm1$~kpc in the distance. The position of the star in the theoretical HR diagram is shown in Fig.~\ref{hr}. The evolutionary tracks from \citet{scha1992}  suggest that the star has just completed H core burning. Its evolutionary mass (obtained placing the star on the theoretical tracks) is $37.5\:M_{\sun}$ (present day), corresponding to an initial mass $M_{{\rm ini}}=42\:M_{\sun}$.

The spectroscopic mass $M_{\ast}=29\pm10\,M_{\sun}$ is (just) consistent within the errors with the present-day evolutionary mass. In many cases, supergiants show a discrepancy between spectroscopic and evolutionary masses \citep{herre92,herrero07}. In addition, the evolutionary mass compares well with those obtained for Ia supergiants in the B0--1 range \citep[e.g.][]{crowther,trundle04}. We must note that the distance to the source may be somewhat higher, allowing for a moderately higher spectroscopic mass. On the other hand, the use of evolutionary tracks that take initial rotation into account \citep[e.g.][]{ekstrom} would result in a somewhat lower evolutionary mass. In conclusion, all the data available are consistent with a present-day mass $\approx35\:M_{\sun}$. We stress that this is an evolutionary mass, and thus directly comparable to other evolutionary masses \citep[e.g.][]{crowther}. The dynamical masses of some HMXBs are smaller than the masses corresponding to their spectral types \citep[e.g.][]{vandermeer}.

\begin{table}
\caption{Chemical abundances resulting from \textsc{fastwind} analyses.\label{table:abundances}}
\smallskip
\centering
\begin{tabular}{lc}
\hline
\hline
\noalign{\smallskip}
Species&log(X/H)\\
\hline
\noalign{\smallskip}
Si&7.73 $\pm$ 0.17\\
O&8.73 $\pm$ 0.20\\
N&8.42 $\pm$ 0.10\\
C&7.74 $\pm$ 0.44\\
Mg&7.60 $\pm$ 0.25\\
\hline
\end{tabular}
\end{table}

\subsection{Light curve}

A long-term photometric lightcurve is available for HD~306414 from the All Sky Automated Survey ASAS-3 photometric catalogue \citep{poj97}. The catalogue contains $V$-band observations of HD~306414 between HJD~2451880 and 2455170, with a total of 629 photometric points.

The lightcurve shows no clear evidence of orbital variability. We searched for possible periodicities using different algorithms (Lomb-Scargle periodograms, phase-dispersion minimisation, {\sc clean}) available within the {\it Starlink} package {\sc period} \citep{dhillon}, without finding any statistically significant result. Next, we folded the photometric data on the 164.6-d period derived from the X-ray observations. Again, we did not find any significant modulation. Short-term variability is present throughout the whole timespan covered by the observations, but it is likely consistent with the photometric errors, typically 0.04~mag, but occasionally as high as 0.08~mag. The average magnitude remains constant over $\sim$9~yr of observations, with a standard deviation of 0.04~mag.

\subsection{Radial velocity curve}
\label{sec:radial}

The strong wind of an early B supergiant may affect the shape and centroid of \ion{H}{i} and \ion{He}{i} lines. For example, \citet{vandermeer} studied the radial velocity curves of the optical components in three high-mass X-ray binaries. These objects, with spectral types between O8 and B0, exhibit only lines of \ion{H}{i}, \ion{He}{i} and \ion{He}{ii} (apart from the \ion{Si}{iv}~4089\,\AA\ line). \citet{vandermeer} carried out measurements of each individual line, finding differences of a few percent between lines.
Fortunately, given its later spectral type, HD~306414 presents a large number of lines corresponding to ionised metals. Such lines are formed in deep photospheric layers, and therefore are not affected by the wind. We performed preliminary analysis of individual spectral lines to decide which set of lines would be most suitable for measuring radial velocities. When individual metallic lines are measured, the radial velocities within a given spectrum are consistent, with differences smaller than $4\,{\rm km}\,{\rm s}^{-1}$. As in previous analyses of similar stars, we found that the best behaviour corresponds to the Si\,{\sc iii} triplet, which presents very small dispersion between the values for the individual lines. However, cross-correlation techniques are well known to produce more accurate velocities, even for early-type stars \citep{liu88}, as they make use of all the spectral information \citep[e.g.][]{hill93}. 

We used the cross-correlation technique, implemented in a python program developed by ourselves, to determine radial velocities. Every cross-correlation function (CCF) was subject to apodization and a Gaussian function was fitted to the CCF by using the method of least squares. We chose as a template the synthetic spectrum used to fit the observed spectrum in the previous subsection. We cross-correlated every individual spectrum using the 4540--4660\,\AA\ range (marked in Fig.~\ref{spec}). There are several strong reasons to choose this region: firstly, it contains the \ion{Si}{iii} triplet, together with a large number of \ion{O}{ii} and \ion{N}{ii} lines; secondly, it does not include any \ion{He}{i} or \ion{H}{i} lines that could be affected by the wind; finally, it does not include any significant atmospheric or interstellar feature that would not participate of the stellar motion. The resulting velocities are shown in Table~\ref{table:rv}, after application of the heliocentric correction. A systemic velocity of $+2.8\:{\rm km}\,{\rm s}^{-1}$, which was applied to the synthetic spectrum, is not included. Phase zero was chosen to correspond to the first radial velocity sorted by date, as we did not want to force the time of periastron passage. The formal error in each individual measurement is $0.9\:{\rm km}\,{\rm s}^{-1}$.

\begin{table}[]
\caption{Radial velocities corrected to the heliocentric restframe and sorted by phase (with a systemic shift of $+2.8\:{\rm km}\,{\rm s}^{-1}$ not applied). The last column shows the residuals for the fit to the orbit shown in Fig.~\ref{sbop} and discussed in Sect.~\ref{orbit}.\label{table:rv}}
\smallskip
\begin{center}
\begin{tabular}{lcrr}
\hline
\hline
\noalign{\smallskip}
Number&Phase&$v_{{\rm rad}}$ (km\,s$^{-1}$)& O-C (km\,s$^{-1}$)\\
\noalign{\smallskip}
\hline
\noalign{\smallskip}

1    &   0.00000  &   $-$6.42    & $-$6.61 \\
2    &   0.04806  &   $-$0.85    & $-$1.70 \\
3    &   0.07255  &    2.78      &  1.56   \\
12   &   0.08557  &   $-$9.46    & $-$11.00\\
13   &   0.09663  &   $-$3.10    & $-$4.83 \\
4    &   0.11563  &    7.64      & 5.67    \\
5    &   0.15761  &   12.23      & 9.39    \\
6    &   0.27233  &   10.49      & 3.62    \\
7    &   0.38911  &   15.01      & 0.20    \\
14   &   0.41835  &   $-$4.15    & $-$3.78 \\
8    &   0.42447  &   $-$6.49    & $-$6.10 \\
9    &   0.43589  &   $-$5.53    & $-$3.65 \\
10   &   0.45430  &    5.54      & 8.64    \\
11   &   0.45436  &    4.97      & 8.07    \\

\noalign{\smallskip}
\hline
\end{tabular}
\end{center}
\end{table}

The persistence of recurrent X-ray outbursts every 164.6~d strongly suggests that this is the orbital period  \citep{rom09}. Our radial velocity data cannot be used to verify the period, as the 2006--2007 data cover only 75 days (i.e. less than half the periodicity), while the 2009 spectra correspond to the same phases when folded on this period. The periodicity in the X-rays has, however, been observed between 2003 and 2011 \citep{romano11} and almost certainly represents the orbital period. In spite of this, when we fold the data on this period, the resulting radial velocity curve is not consistent with expectations. For instance, measurements $\sharp3$ and $\sharp12$ were taken at almost the same phase (though on different dates), but show a large difference in radial velocity  ($>14\:{\rm km}\,{\rm s}^{-1}$). This effect is not due to the measurement technique, as illustrated by measurements $\sharp10$ and $\sharp11$. These two spectra, taken on the same night, result in measurements identical within the errors, demonstrating that the cross-correlation technique is as accurate as the formal errors indicate.

\begin{figure}[!h]
\begin{center}
\resizebox{\columnwidth}{!}{\includegraphics[clip]{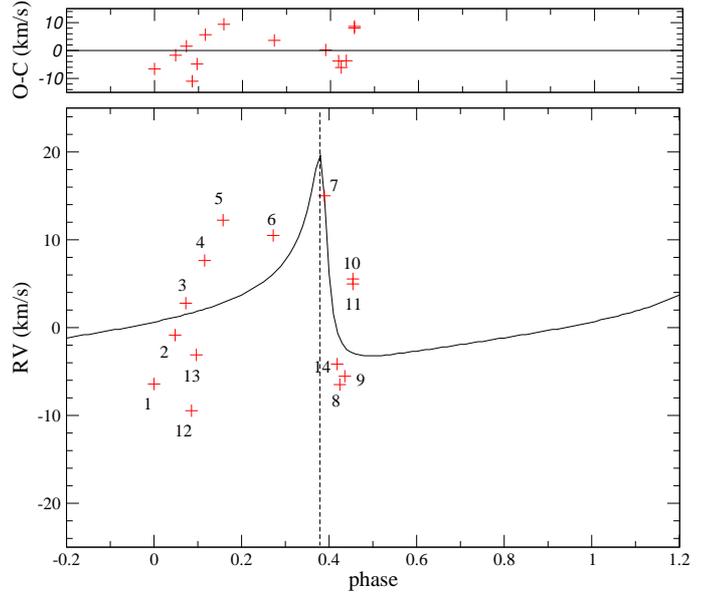}}
\end{center}
\caption{Radial velocity curve folded on the 164.6-d period derived from the X-ray observations, with phase zero chosen to be the time of the first observation. The dashed line marks the time of X-ray peaks. The thick line represents a typical radial velocity curve for an eccentric orbit with the periastron phase at this time (see Section~\ref{orbit}). The numbers identify the observations, as in Table~\ref{table:log}. \label{sbop}}
\end{figure}

Moreover, the folded radial velocity curve (Fig.~\ref{sbop}) does not show a shape compatible with orbital modulation. The dispersion of the measurements is clearly dominated by two large excursions, happening around phase 0.1 and around phase 0.4. We are thus forced to conclude that the radial velocity changes in the spectrum of HD~306414 are not directly reflecting the dynamical motion of the system, but are caused at least in part by some other physical effect. We do not believe these changes to be due to variations in line profiles associated with the stellar wind, because they are measured in high-ionisation lines (such as the \ion{Si}{iii} triplet), but also seen in lines like \ion{He}{i}~4471\,\AA, more likely to be affected by the wind.

\begin{figure}[h]
\resizebox{\columnwidth}{!}{\includegraphics[clip]{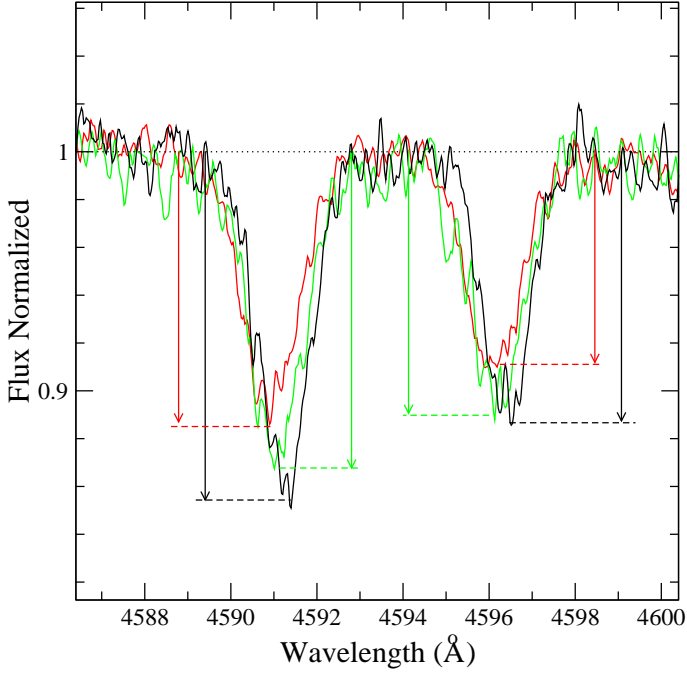}}
\caption{Variability in the line profiles of two of the features used for the cross correlation, \ion{O}{ii}~4591 and~4596\,\AA. The red line is spectrum~1, the green line is spectrum~3, and the black line is spectrum~5. The variations in line depth (indicated by the arrows) are typical of pulsating supergiants. \label{pulse}}
\end{figure}

A similar effect, with radial velocity variations not reflecting binary motion, is observed in the HMXB GX301$-$2, which contains a very luminous B1 hypergiant \citep{kaper2006}. Possible explanations are tidal deformation (unlikely in such a wide system as IGR~J11215$-$5952, except close to periastron) or pulsations \citep[cf.][]{kerk1995,quaint03}. To investigate the origin of the variations, we plotted the evolution of several metallic lines, which are expected to be produced deep in the photosphere and not affected by the stellar wind, finding very complex variability. In addition to changes in the radial velocity of the centroids, the lines display changes in shape and depth. An example is shown in Fig.~\ref{pulse}, but all metallic lines display similar behaviour. Such changes have been observed in other luminous B-type supergiants, and are generally attributed to pulsations \citep[e.g.][]{ritchie09}.

\begin{figure}[h]
\resizebox{\columnwidth}{!}{\includegraphics[clip]{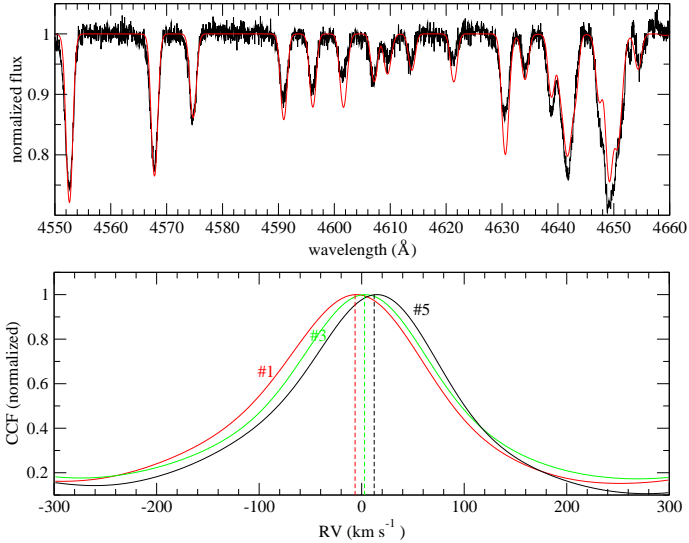}}
\caption{{\bf Top panel: } Comparison of the model spectrum with spectrum~1 in the region used for the cross-correlation. {\bf Bottom panel: }Peak of the cross correlation functions for three spectra. The spectra chosen are the same ones for which a small fraction is depicted in Fig.~\ref{pulse}. The colour coding is the same as in Fig.~\ref{pulse}. \label{ccfs}}
\end{figure}

Given the important changes in shape of the lines used for the cross-correlation, one could worry about the accuracy of the radial velocities measured. In Fig.~\ref{ccfs}, we plot the cross-correlation functions for the three spectra that were shown in Fig.~\ref{pulse}. In spite of the large differences in the shapes of the metallic lines, the shape of the cross-correlation function is not affected, and therefore we conclude that the relative accuracy of the radial velocity determinations is as good as their formal errors.

\begin{figure}[h]
\resizebox{\columnwidth}{!}{\includegraphics[clip]{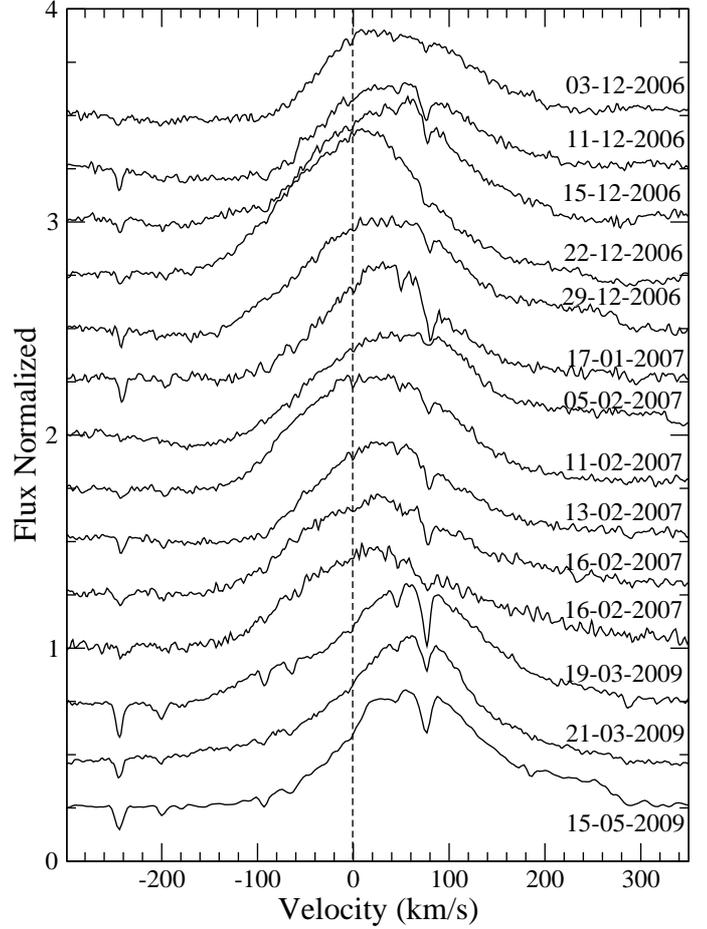}}
\caption{H$\alpha$ profiles normalised and displayed in temporal order (day-month-year). The variable sharp features are atmospheric. \label{alfa}}
\end{figure}

The H$\alpha$ line, the best mass loss tracer in the optical range, also shows strong variability in shape and radial velocity (as defined by the centroid of the emission component). The changes in radial velocity do not correlate with those of the photospheric lines. This lack of correlation is not unexpected, as the emission line should be probing the stellar wind. The variations detected in the H$\alpha$ line are likely due to structure
in the stellar wind. In some high-mass X-ray binaries with close orbits, the wind structure is affected by the X-rays, and some absorption components are attributed to an accretion wake \citep[e.g.][]{kaper94}. Such structures are very unlikely to form in IGR~J11215$-$5952, because X-rays are only emitted over a very small part of the orbit.

In spite of this, as seen in Fig.~\ref{alfa}, very strong changes in the shape of H$\alpha$ took place close to the time of the X-ray outburst covered by our observations, which started on February 7th and had its peak on February 9th \citep{rom09}. The H$\alpha$ profile on February 5th clearly shows a much deeper absorption trough than any other spectrum. Meanwhile, the centroid of the emission features migrates redwards. In contrast, on February 11th the emission feature had moved bluewards, showing a very large excursion in $v_{{\rm rad}}$. By February 13th, the line seems back to its typical profile. We must note, however, that a very strong bluewards excursion is also seen in the December 22nd spectrum, seven weeks (i.e. almost one third of the orbital period) before the outburst, when no X-ray emission was expected from the source. While these excursions in the radial velocity of H$\alpha$ are taking place, the radial velocity of the photospheric lines does not follow the same behaviour.

\section{Discussion}
\label{sec:discu}

IGR~J11215$-$5952 is a peculiar X-ray source, associated with the class of SFXTs. A neutron star, detected as a transient $P_{{\rm S}}=187\:{\rm s}$ pulsar, orbits the B0.5\,Ia supergiant HD~306414. The X-ray source is only detected for short timespans ($<10$~d), but reappears consistently every 164.6~days, strongly suggesting that this is the orbital period \citep{romano11}. Because of the very short duty cycle, a very high eccentricity is assumed. Two basic ideas have so far been proposed to explain its behaviour. \citet{sido2007} speculated on the presence of an equatorial disk around the supergiant, with the outbursts being due to crossings of the neutron star through this disk. \citet{negue2008} tried to fit IGR~J11215$-$5952 within a general picture of supergiant X-ray binaries, suggesting that accretion becomes inefficient when the neutron star is more than $\sim 3\,R_{*}$ away from the supergiant, as a consequence of increasing clumpiness in the stellar wind.

The presence of a large disk around the supergiant seems to be ruled out by the optical observations, as there is no evidence of such a structure in the optical spectra. The H$\alpha$ line, the best mass-loss tracer in the optical range, shows a morphology and variability typical of early-B supergiants (Fig.~\ref{alfa}), and can be fit by the model atmosphere, which assumes spherical mass loss. A small disk can perhaps be accommodated, by assuming that it is outshone by the supergiant, but then the outbursts must happen when the neutron star is within the region with a strong stellar wind, and the two alternatives become almost indistinguishable. Another closely related possibility is an equatorial enhancement or focusing of the wind. This possibility has been
proposed in some similar systems with very luminous companions \citep[e.g.][]{zand98}, though no physical mechanism has been proposed to cause this enhancement.

On the other hand, accretion from a spherical wind, even if clumpy, seems unable to reproduce the characteristics of the X-ray outbursts, such as the very short duration and strong peak \citep{rom09,karino10}. An abrupt transition between a region where high accretion rates can be achieved and the rest of the orbit, where accretion is suppressed, would be required.

The optical spectra show a strong change in the morphology of H$\alpha$ around the time of the February 2007 periastron passage. As we also see another important change seven weeks before periastron, a direct connection between the changes in profile and the periastron passage cannot be claimed. The spectral morphology of HD~306414 indicates a very luminous supergiant, and hence a strong and highly variable mass-loss is expected. The conclusion that changes in radial velocity do not seem to be dominated by orbital motion, but by some other phenomenon, most likely stellar pulsation, adds complication to the stellar mass loss. The possibility of enhanced mass loss close to periastron, though, is not ruled out, and perhaps is even suggested by the observations.

\subsection{Stellar pulsations}

The sort of line-profile variability displayed in Fig.~\ref{pulse} is present in all the absorption lines in the spectrum of HD~306414. Similar variability is generally interpreted as a reflection of stellar pulsations. Oscillations in main-sequence and not-very-evolved B-type stars have been known for years \citep{aerts06}. Recent evidence also points to widespread oscillations in B-type supergiants. Both g- and p-modes were detected in HD~163899 \citep[B1\,Ib,][]{saio}, while photometric variability strongly suggests that a very high fraction of B supergiants present opacity-driven gravity-mode oscillations \citep{lefever07}. In addition, there is growing evidence suggesting that the extra line broadening in OB supergiants known as macroturbulence could be caused by photospheric line-profile variations \citep{aerts09,sergio13}.

Oscillations have also been reported in GP~Vel, the B0.5\,Ib optical counterpart to the X-ray pulsar Vela X-1. When analysing its radial velocity curve, \citet{kerk1995} detected substantial deviations from the curve expected for pure Keplerian motion. Large changes in the shape of line profiles were also observed. Later, \citet{quaint03} showed that, after subtracting the best orbital fit, the residual radial velocity excursions seemed to be modulated at multiples of the orbital frequency. This led \citet{quaint03} to suggest that the oscillations were tidally induced by the companion.

\citet{koen12} showed that tidal interactions between the neutron star and supergiant in the slightly eccentric Vela X-1 system can indeed induce surface motions on the supergiant companion that lead to strongly variable profiles for the photospheric lines. This in turn causes the measured radial velocity curve to deviate significantly from that expected from Keplerian motion. In all simulations run by \citet{koen12}, the net effect was to give a radial velocity amplitude higher than expected from pure Keplerian motion, which in turn would lead to an overestimation of the neutron star mass, as pointed out by \citet{quaint03}. Similar effects could also occur in HD~306414, which has an even more eccentric orbit, but without detailed modelling of the type carried out by \citet{koen12}, such an assertion remains speculative.

The radial velocity excursions measured in HD~306414 are not much more extreme than those seen in GP~Vel, which can reach $\approx12\:{\rm km}\,{\rm s}^{-1}$  \citep{quaint03}, and are also compatible with the highest values observed in (apparently) isolated supergiants \citep{sergio10}. It is unclear if the neutron star companion may play a role in exciting the oscillations. The orbit of Vela X-1 is much tighter ($P_{{\rm orb}}=9.0\:{\rm d}$), but the eccentricity is almost certainly much smaller ($e=0.09$ in Vela X-1). In a wide, eccentric orbit, the regular passage of the neutron star through a close periastron could provide a strong resonance effect, exciting non-radial oscillations \citep{witte}. In any case, we must stress that the presence of a companion does not seem necessary to excite strong pulsation. Similarly wide excursions have been observed in the extreme B supergiant HD~50064, which could be related to luminous blue variables \citep{aerts10}. They seem to be modulated with a 37~d period, also detected in the photometry, which is interpreted as a radial oscillation mode. Strong changes in the depth of photospheric lines are also seen \citep{aerts10}.

Pulsations are generally revealed by periodic photometric variability. However, the typical amplitudes of these variations are only a few hundredths of a magnitude \citep[typically, $\sim$0.04;][]{lefever07}. Variability of this amplitude would not be detectable with the accuracy and sampling of the ASAS photometry. We also checked if the presence of pulsations may have an effect on the stellar parameters derived. We carried out the same {\sc fastwind} analysis for all the spectra available, finding slightly different physical parameters between spectra, but always compatible with the values reported within the error bars quoted.

\subsection{Stellar and orbital parameters}
\label{orbit}

Even if stellar pulsations are the main cause of the radial velocity variations, it is highly unlikely that they are the {\it only} cause. Our spectroscopic analysis of HD~306414 resulted in values for the mass compatible with a current mass $M_{*}\approx35\,M_{\sun}$. Naturally, a neutron star, with a mass $M_{{\rm X}}\approx 1.4\,M_{\sun}$ in a wide orbit is unlikely to induce large Doppler shifts in such a massive supergiant. On the other hand, the short X-ray outbursts detected every 164.4~d argue for a highly eccentric orbit, suggesting that the neutron star should come sufficiently close to the supergiant to induce some measurable radial velocity shifts, unless the line of sight is almost perpendicular to the orbital plane. It is thus sensible to expect an orbital signature to lie hidden below the higher-amplitude variations due to pulsations.

In an attempt to constrain the orbital parameters, we used the SBOP code\fnmsep\footnote{{\tt http://mintaka.sdsu.edu/faculty/etzel/}} \citep{etzel}, which fits single-lined orbits to the observed radial velocities of a spectroscopic binary using one of several optimisation schemes based on the Lehmann-Filhes differential correction procedure, to investigate which sort of orbital solutions are compatible with the observed radial velocity curve. SBOP requires a preliminary knowledge of some parameters to produce an accurate fit. If we fix the orbital period to the X-ray outburst recurrence time, the code will invariably locate the periastron at the phase corresponding to the large excursion in radial velocity around 2006 December 22nd. The sudden increase in radial velocity in a few days is interpreted by the code as periastron passage in a high-eccentricity orbit. As it is difficult to explain the short X-ray outbursts observed if they happen at orbital phase $\sim0.3$, we assume that this abrupt change in radial velocity is not due to orbital motion.

Therefore we also forced the time of periastron to be constrained between MJD~54136 and 54144, i.e. within four days of the peak of the X-ray outburst that happened on 2007 February 9th and fixed the zero time of ephemerides to the time of the first observation. We run SBOP fixing the eccentricity and letting the other orbital parameters converge.  We tried different values of the eccentricity, varying in steps of 0.05, and checked the standard deviation of the fit for each value. Orbits with moderate to large eccentricities can be fitted to the radial velocity points with similar standard deviations, though none of them results in a credible fit. From examination of a large number of fits, we conclude that the semi-amplitude for the optical component must be $K_{{\rm opt}}\la11\pm6\:{\rm km}\,{\rm s}^{-1}$ so that the orbital variations are not seen.

The large changes in radial velocity observed around the time of periastron allow for rather high orbital eccentricity. If we assume $K_{{\rm opt}}\approx11\:{\rm km}\,{\rm s}^{-1}$, we can find solutions with high-eccentricity that result in relatively low standard deviations (as compared to other solutions). As an example, Figure~\ref{sbop} shows the radial velocity curve of an orbit with $K_{{\rm opt}}=10.4\:{\rm km}\,{\rm s}^{-1}$ and $e=0.8$ that could lie hidden below the pulsations. As an illustration, Table~\ref{table:rv} includes in the last column the residuals for this fit. Even though they are unacceptably high, no other fit results in a lower standard deviation. For such high eccentricity orbits, assuming $M_{*}=35\,M_{\sun}$ and a standard neutron star mass $M_{{\rm X}}=1.4\,M_{\sun}$, the expected  $K_{{\rm opt}}$ is always $\approx10\:{\rm km}\,{\rm s}^{-1}$ for $i=90\degr$. It is therefore not surprising that we fail to detect the orbital motion in the presence of strong pulsations.

\subsection{Accretion mechanisms}
The orbital parameters suggested by the X-ray behaviour of IGR~J11215$-$5952 are very different from those of other SFXTs (or any other supergiant X-ray binary), as both the orbital period and eccentricity appear rather high, and seem more typical of long-period Be/X-ray binaries \citep{okneg01}. Supergiant systems like IGR~J16465$-$4507 or SAX~J1818.6$-$1703 have $P_{{\rm orb}}\approx30$~d, and moderate to high eccentricities are deduced from the modulation in their X-ray flux \citep[see, e.g.][]{dclark10}.  The peculiar supergiant X-ray binary GX301$-$2 has $P_{{\rm orb}} = 41.498$~d and $e = 0.462$ \citep{koh97}. Though its X-ray flux is strongly modulated and presents a strong peak, emission is detected throughout the orbit. \citet{leahy08} find that the X-ray lightcurve can be reproduced if the B1\,Ia$^{+}$ hypergiant companion loses mass through a tidal stream in addition to the strong stellar wind.

Even though the mass donor in GX301$-$2 is more evolved and perhaps more massive (and hence more luminous) than HD~306414 \citep{kaper2006}, the similarities between the two systems are strong. It is easy to envisage IGR~J11215$-$5952 evolving into a system very similar to GX301$-$2 as stellar evolution turns HD~306414 into a later-type B hypergiant (with a slower and denser wind) and tidal circularisation reduces the eccentricity. The accretion mechanism may also be similar. Given the high mass of the supergiant, the neutron star may come very close to its surface without giving rise to changes in the radial velocity higher than those allowed by our observations. Indeed, periastron distances $\approx 2\:R_{*}$ are compatible with eccentricities $\approx0.8$. Such a close distance may lead to localised mass loss through the inner Lagrangian point in the form of a transient tidal stream. This scenario is fully consistent with the large variations in the shape of H$\alpha$ observed close to the periastron passage and may explain the strong perturbing influence of the neutron star on the supergiant companion in spite of the large average distance.

The X-ray outbursts would then be due to the accretion of stellar wind from regions close to the stellar atmosphere coupled with a transient tidal stream that may even permit the formation of a transient accretion disk. The presence of such transient structure may be tested by future high-sensitivity missions via the study of spin evolution during the short X-ray outburst. With the high-eccentricity orbit suggested by the shape of the outbursts, the neutron star will quickly move to distances $d>3\,R_{*}$, leading to the disappearance of the tidal stream (and putative accretion disk) and hence the abrupt decrease in luminosity. The residual emission that is sometimes seen up to 6--8 days after the peak \citep{rom09} may then be due to the low-density stellar wind at high distances.

Based on the orbital parameters of IGR~J11215$-$5952, \citet{liu11} suggested that the system might have gone through a phase as a Be/X-ray binary in the past. The high mass that we derive for HD~306414 makes this possibility very unlikely. With an initial mass $\ga35\:M_{\sun}$, HD~306414 should have had a spectral type around O5\,V when it was on the main sequence \citep{martins}. No Oe star with such an early spectral type is known \citep{neg04}. Moreover, in the accretion scenario presented here, a transient accretion disk may possibly form near periastron, allowing the transfer of angular momentum to the neutron star and thus invalidating the assumption of pure wind accretion used by \citet{liu11}.

\section{Conclusions}

We used high-resolution spectroscopy of HD~306414, the optical counterpart to IGR~J11215$-$5952, to determine its astrophysical parameters and search for orbital modulation. We find that HD~306414 is a luminous B0.5\,Ia supergiant, at a distance not smaller (and perhaps slightly higher) than 7~kpc. Its present-day mass is $\approx35\,M_{\sun}$, a typical value for the spectral type. Its chemical composition is also typical of the spectral type, and reveals little chemical evolution, suggesting that the star is just ending core hydrogen burning.

The star presents moderately strong variations in radial velocity, but these changes take the form of large excursions on a timescale of a few days and do not seem to reflect orbital motion. Very significant changes in shape and depth of all photospheric lines are taken as indications of pulsations, providing a likely origin for the radial velocity excursions. Variations in radial velocity reflecting the orbital motion are not evident, but we checked that the signature of a wide, highly-eccentric orbit, as suggested by the X-ray behaviour, can be effectively masked by the effects of pulsations.

In view of these characteristics, we suspect that IGR~J11215$-$5952 is not a typical SFXT, but rather a system in which flares are driven by the close approach of the neutron star to the companion during periastron, very likely resulting in localised mass loss from the outer layers of the supergiant and the formation of a transient accretion disk. The close passage may be (partially) responsible for the excitation of the pulsational modes. As the supergiant expands, IGR~J11215$-$5952 will probably turn into a system very similar to GX301$-$2, which contains a neutron star in an eccentric orbit around a B1.5\,Ia$^{+}$ hypergiant.

\begin{acknowledgements}

We are very thankful to Jes\'us Ma\'{\i}z-Apell\'aniz for deriving the extinction law to HD~306414 with the latest version of the {\sc chorizos} code. We also thank Sergio Sim\'on-D\'{\i}az for very useful discussions on stellar pulsations in OB stars, and the anonymous referee for useful suggestions.

This research is partially supported by the Spanish Ministerio de
Ciencia e Innovaci\'on (MICINN) under
grants AYA2010-21697-C05-04/05, AYA2012-39364-C02-01/02, and CSD2006-70 (First Science with the GTC), and by the Generalitat Valenciana under grant ACOMP/2012/134.

 This publication
makes use of data products from the Two Micron All
Sky Survey, which is a joint project of the University of
Massachusetts and the Infrared Processing and Analysis
Center/California Institute of Technology, funded by the National
Aeronautics and Space Administration and the National Science
Foundation.

\end{acknowledgements}

{}
\end{document}